\documentclass[12pt]{article}
\usepackage[latin1]{inputenc}
\usepackage{lipsum}
\usepackage{amsmath}
\usepackage{color}
\usepackage{amsfonts}
\usepackage[mathscr]{euscript}
\usepackage{graphicx}
\usepackage{geometry}
\usepackage{amssymb,epsfig}
\usepackage{subfigure}
\usepackage{comment}
\usepackage{tabularx}
\usepackage{bm}
\usepackage{euscript}
\usepackage[dvipsnames,table]{xcolor}
\usepackage{exscale}
\usepackage{amsbsy}
\usepackage{textcomp}
\usepackage{hyperref}
\usepackage{slashed}
\usepackage{authblk}
\usepackage{tabularx}
\usepackage{color}
\usepackage{exscale}
\usepackage{tcolorbox}
\usepackage{doi}
\usepackage{tensor}
\usepackage{wrapfig}
\usepackage[font=footnotesize,labelfont=bf]{caption}
\usepackage{ulem} 
\usepackage{makecell}

\def\ba#1\ea{\begin{align}#1\end{align}}
\def\bg#1\eg{\begin{gather}#1\end{gather}}
\def\bm#1\em{\begin{multline}#1\end{multline}}
\def\bmd#1\emd{\begin{multlined}#1\end{multlined}}

\newcommand{\be}{\begin{equation}}
	\newcommand{\ee}{\end{equation}}
\newcommand{\bea}{\begin{eqnarray}}
	\newcommand{\eea}{\end{eqnarray}}

\newcommand{\matleft}{\left(\begin{array}}
	\newcommand{\matright}{\end{array}\right)}

\newcommand{\ea}{\end{eqnarray} }

\newcommand{\bpm}{\begin{pmatrix}}
\newcommand{\epm}{\end{pmatrix}}
\newcommand{\bmm}{\begin{matrix}}
\newcommand{\emm}{\end{matrix}}

\newcommand{\la}{\label}
\newcommand{\p}{\partial}

\definecolor{green2}{RGB}{82,175,126}
\definecolor{orange2}{RGB}{252,141,98}
\definecolor{blue2}{RGB}{141,160,203}
\usepackage[numbers,sort&compress]{natbib}
\def\simge{
	\mathrel{\rlap{\raise 0.511ex 
			\hbox{$>$}}{\lower 0.511ex \hbox{$\sim$}}}}

\def\simle{
	\mathrel{\rlap{\raise 0.511ex 
			\hbox{$<$}}{\lower 0.511ex \hbox{$\sim$}}}}

\makeatletter
\renewcommand\section{\@startsection {section}{1}{\z@}%
	{-3.5ex \@plus -1ex \@minus -.2ex}
	{2.3ex \@plus.2ex}%
	{\normalfont\large\bfseries}}
\renewcommand\subsection{\@startsection{subsection}{2}{\z@}%
	{-3.25ex\@plus -1ex \@minus -.2ex}%
	{1.5ex \@plus .2ex}%
	{\normalfont\bfseries}}
\renewcommand\subsubsection{\@startsection{subsubsection}{3}{\z@}%
	{-3.25ex\@plus -1ex \@minus -.2ex}%
	{1.5ex \@plus .2ex}%
	{\normalfont\itshape}}
\makeatother

\def\pplogo{\vbox{\kern-\headheight\kern -29pt
		\halign{##&##\hfil\cr&{\ppnumber}\cr\rule{0pt}{2.5ex}&\ppdate\cr}}}
\makeatletter
\def\ps@firstpage{\ps@empty \def\@oddhead{\hss\pplogo}%
	\let\@evenhead\@oddhead 
}
\hypersetup{
	unicode=false,          
	pdftoolbar=true,        
	pdfmenubar=true,        
	pdffitwindow=false,     
	pdfstartview={FitH},    
	pdftitle={Korteweg de-Vries Dynamics at the Edge of Laughlin State},    
	pdfauthor={},     
	pdfsubject={Subject},   
	pdfcreator={},   
	pdfproducer={}, 
	pdfkeywords={keyword1} {key2} {key3}, 
	pdfnewwindow=true,      
	colorlinks=true,       
	linkcolor=OliveGreen, 
	citecolor=NavyBlue,        
	filecolor=magneta,      
	urlcolor=cyan           
}

\numberwithin{equation}{section}

\newcommand*\samethanks[1][\value{footnote}]{\footnotemark}

\newcommand\beal{\begin{equation}\begin{aligned}}
		\newcommand\eeal{\end{aligned}\end{equation}}

\begin{document}

\normalem

\setcounter{page}0
\def\ppnumber{\vbox{\baselineskip14pt
}}

\def\ppdate{
} 
\date{}

\title{\Large\bf Korteweg de-Vries Dynamics at the Edge of Laughlin State}
\author{Gustavo M. Monteiro$^1$, Sriram Ganeshan$^{2,3}$}
\affil{\it\small $^1$ Department of Physics and Astronomy, College of Staten Island, CUNY, Staten Island, NY 10314, USA}
\affil{\it\small $^2$ Department of Physics, City College, City University of New York, New York, NY 10031, USA}
\affil{\it\small $^3$ CUNY Graduate Center, New York, NY 10031}
\maketitle\thispagestyle{firstpage}
\begin{abstract}

In this work, we show that the edge dynamics of the Laughlin state in the weakly nonlinear regime is governed by the Korteweg-de Vries (KdV) equation. Our starting point is the Chern-Simons-Ginzburg-Landau theory in the lower half-plane, where the effective edge dynamics are encoded in anomaly-compatible boundary conditions. The saddle point bulk dynamics and the corresponding boundary conditions of this action can be reformulated as two-dimensional compressible fluid dynamic equations, subject to a quantum Hall constraint that links the superfluid vorticity to its density fluctuations. The boundary conditions in this hydrodynamic framework consist of no-penetration and no-stress conditions. We then apply the method of multiple scales to this hydrodynamic system and derive the KdV equation for the edge dynamics in the weakly nonlinear regime. By employing the Hamiltonian framework for the KdV equation, we show that we can recover the chiral Luttinger liquid theory in the linearized regime and provide a pathway for canonically quantizing the edge dynamics in the weakly non-linear limit.
\end{abstract}

\pagebreak
{
\hypersetup{linkcolor=black}
\tableofcontents
}
\pagebreak


\section{Introduction}

Laughlin states are many-body quantum states that emerge in a two-dimensional electron gas under strong magnetic fields and represent the simplest examples of the fractional quantum Hall effect~\cite{laughlin1983anomalous, tsui1982two}. These states are characterized by their highly correlated nature, which leads to fractionally charged excitations with anyonic statistics. Edge dynamics in the Laughlin state are frequently modeled using chiral Luttinger Liquid ($\chi$LL) theory, which describes the propagation of gapless edge excitations along the system's boundaries~\cite{wen1990compressibility}. These chiral edge states are crucial for robust transport associated with the quantized Hall conductance. Their behavior can be probed experimentally through edge tunneling experiments and shot noise measurements~\cite{wen1991edge}. Analytical techniques and numerical simulations have investigated the interplay between bulk and edge physics in the Laughlin state. Analytical methods typically involve conformal field theory (CFT), particularly $\chi$LL theory, to describe edge excitations. This is related to a bulk $U(1)$ Chern-Simons theory that quantifies fractional charges and their statistics~\cite{wen1990compressibility, wenbook}. Numerical techniques, such as exact diagonalization based on the Laughlin wave function and density matrix renormalization group methods, provide a detailed understanding of ground state properties and excitations within a confining potential, effectively capturing the bulk-edge correspondence~\cite{yoshioka1983ground, laughlin1983quantized, feiguin2008density, li2008entanglement, chandran2011bulk}. These approaches capture several qualitative aspects of the quantum Hall edge transport observed in experiments.

Among several remarkable experiments on quantum Hall systems, two, to the best of our knowledge, stand out for providing direct evidence of non-linear effects at the edge. The first, an earlier experiment on GaAs, reported dissipative scaling on the quantum Hall edge, describing the behavior as ``somewhere between linear and quadratic"~\cite{talyanskii1994experimental}. This phenomenon is attributed to the interplay between non-linearities and diffusion effects, as discussed in our recent work~\cite{monteiro2023kardar}. The second experiment directly observed shock waves at the quantum Hall edge~\cite{zhitenev1995linear}. These findings suggest interesting non-linear dynamics and may indicate the importance of interaction effects near the edge. Both analytical and numerical efforts have been made to quantify these non-linear dynamics in the quantum Hall edge. Wexler and Dorsey studied the vortex patch dynamics in a classical plasma, showing that the resulting contour dynamics is described by the modified KdV (mKdV) equation using the Direct Interaction Approximation (DIA)~\cite{wexler1999contour}. Bettelheim et al. proposed the integrable Benjamin-Ono (BO) equation to model non-linear edge dynamics, with the BO solitons carrying one ``quantum" of anyonic charge~\cite{bettelheim2006nonlinear}. In this scenario, a Cauchy initial value problem for an electron wave packet evolves into a train of BO solitons, each carrying anyonic charge. A conformal field theory (CFT)-based analysis by Fern, Bondesan, and Simon~\cite{fern2018effective} showed that the non-local dispersion of the BO equation is incompatible with the CFT structure of the quantum Hall edge. Their work also suggests that KdV dynamics is consistent with the edge CFT, making it a potential extension to the $\chi$LL theory~\cite{fern2018effective}.

 More recently, numerical works by Nardin and Carusotto on edge excitations in IQH and FQH systems provided evidence of KdV dispersion in edge dynamics~\cite{nardin2023linear}. The possibility of integrable non-linear dynamics at the quantum Hall edge suggests the need for a systematic approach to derive the KdV equation for the Laughlin state. Historically, Korteweg and de Vries first derived the KdV equation in a shallow water context using the method of multiple scales, in a regime where the wavelength of surface waves is much larger than the depth of the channel~\cite{korteweg1895xli}. A natural question arises: Can a similar approach, in the spirit of the original Korteweg-de Vries work, be applied to a ``hydrodynamic" model of the Laughlin state?

The bulk hydrodynamic equations of the Laughlin state can be derived from the Chern-Simons-Ginzburg-Landau (CSGL) theory, which describes a charged condensate coupled to a statistical Chern-Simons gauge field~\cite{read1989order, zhang1989effective, zhang1992chern} also known as the composite boson theory. This composite boson approach is complementary to the composite fermion picture, where the Laughlin state is viewed as a collective state of composite particles formed by an electron with two attached flux quanta~\cite{jain2007composite}. Although both the composite boson and fermion approaches are thought to yield similar qualitative descriptions of the Laughlin states~\cite{jain2007composite}, the composite fermion model does not possess a superfluid or traditional hydrodynamic interpretation. 
As our study focuses solely on the Laughlin states and aims to develop a hydrodynamic description of these states, we will concentrate on the composite boson CSGL theory from this point forward. Michael Stone originally provided a superfluid hydrodynamic interpretation of the CSGL action in~\cite{stone1990superfluid}, which was later generalized to include Hall viscosity in~\cite{abanov2013effective}. A hydrodynamic-like system of equations emerges from the saddle-point dynamics of composite bosons when expressed in terms of hydrodynamic equations using the Madelung transformation. These equations are supplemented by an additional constitutive relation, originating from the Chern-Simons term, where fluid vorticity is linked to density fluctuations against a constant background. This additional equation is often called Hall constraint. In our recent work~\cite{monteiro2022topological}, we established a duality between the CSGL action and the fluid dynamical Clebsch action with topological terms, as discussed by Nair~\cite{nair2021topological}. We also derived anomaly-compatible boundary conditions at a hard-wall edge. Furthermore, we showed that the resulting fluid dynamical edge action, with these boundary conditions, is a non-linear generalization of the chiral boson action coupled to matter density. In a follow-up study~\cite{monteiro2023coastal}, we showed that the CSGL hydrodynamic system with hard-wall boundary conditions is closely related to the shallow water equations in ocean dynamics. The key distinction is that the CSGL system includes higher derivative terms, which require two boundary conditions at the hard wall (no penetration and no stress boundary conditions), as opposed to the single no-penetration condition in the shallow water equations. We further identified two chiral modes at the edge, referred to as the Kelvin mode and the chiral boson mode. Contrary to earlier discussions in the literature~\cite{nagaosa1994chern, orgad1996coulomb, orgad1997chern}, we demonstrated that the Kelvin mode cannot be the topological edge mode of the Laughlin state, whereas the chiral boson mode satisfies the bulk-boundary correspondence expected of an edge mode in a topological system.

In this paper, starting from the hydrodynamic formulation of the CSGL theory in the lower half-plane with hard-wall boundary conditions, we perform a weakly non-linear analysis using the method of multiple scales, in the spirit of the Korteweg-de Vries approach. We find that the weakly non-linear dynamics of the condensate density fluctuations, \(\delta n\), lead to the following KdV equation at the hard wall edge:
\begin{equation}
\left[ \partial_t(\delta n) + 2c\ell_B\omega_B\,\partial_x(\delta n) + \frac{c}{2}\ell_B^3\omega_B\,\partial_x^3(\delta n) + c\frac{4c^2-1}{4c^2+1}\,\frac{2\pi\ell_B^2\omega_B }{\nu}\,(\delta n)\,\partial_x(\delta n) \right]\bigg|_{y=0} = 0 \,, \label{eq:KdVintro}
\end{equation}
where \(c\) is the short-range interaction parameter (dimensionless sound velocity), \(\nu\) is the filling factor, \(\ell_B\) is the magnetic length, and \(\omega_B\) is the cyclotron frequency. The first two terms in Eq.~(\ref{eq:KdVintro}) describe the ballistic behavior of the chiral Luttinger liquid, where local electron-electron interactions in the bulk fully determine the group velocity of the edge profile. The group velocity can only depend on the interaction parameter in the absence of a confining potential when hard wall boundary conditions are imposed. The third term accounts for dispersive corrections to the chiral Luttinger liquid model, while the nonlinear term captures the effects of the self-interaction at the edge.

This paper is organized as follows: In Sec.~\ref{sec:review}, we review the composite boson model, or CSGL action, under hard wall boundary conditions. We also discuss the derivation of anomaly-compatible hydrodynamic boundary conditions that align with the chiral dynamics expected at the boundary of a Laughlin state. In Sec.~\ref{sec:edgemodes}, we introduce fast and slow time scales and non-dimensionalize the fields and variables using characteristic length and time scales. We then establish the relationships between these scales to ensure that dispersive and nonlinear effects are of the same order in the slow time scale dynamics. Subsequently, we analyze the linearized dynamics in the ballistic regime and identify the existence of two modes. In Sec.~\ref{sec:CBmode}, we examine the chiral boson mode in detail, determining both its dynamical equation and its edge Hamiltonian. By imposing the existence of a Hamiltonian structure, we derive the Poisson algebra of density fluctuations at the edge. In Sec.~\ref{sec:Kelvin}, we investigate the Kelvin mode and its instability. Finally, we discuss how these nonlinear behaviors can be observed in experiments.

\section{Review of Composite Boson Superfluid Dynamics}
\label{sec:review}

In this section, we begin by reviewing Stone's work in Ref.\cite{stone1990superfluid}, which relates the equations of motion of the CSGL action to a set of fluid dynamic equations, incorporating an additional constitutive relation that links superfluid vorticity to fluctuations in condensate density. Unlike Ref.\cite{stone1990superfluid}, we retain higher-order derivatives of the fields, as the assumption of neglecting them breaks down near a hard-wall boundary, where condensate density can fluctuate significantly over distances comparable to the magnetic length, as demonstrated in~\cite{orgad1996coulomb,orgad1997chern,monteiro2022topological}.

The CSGL action describes a charged bosonic matter field, $\Phi$, coupled to an external electromagnetic field, $A_\alpha$, and a statistical Chern-Simons gauge field, $a_\alpha$. Here, $\alpha$ denotes both temporal and spatial components of the vector fields, specifically $\alpha=0,1,2$. For a domain without boundaries, the action is given by:
\begin{align}
S_{\text{CSGL}}&=\int \left[i\hbar\Phi^\dagger D_t\Phi - \frac{\hbar^2}{2m}|D_i\Phi|^2 - V(|\Phi|^2) + \mu|\Phi|^2 - \frac{\hbar\nu}{4\pi}\epsilon^{\alpha\beta\gamma} a_\alpha\partial_\beta a_\gamma\right]d^3x\,, \la{S-CSGL}
\end{align}
where $\mu$ is the chemical potential, $D_\alpha = \partial_\alpha + ia_\alpha + \frac{e}{\hbar}A_\alpha$ is the covariant derivative and, as mentioned earlier in the introduction, $\nu$ is the filling factor. Here, $e$ and $m$ represent the charge and effective mass of the underlying  quasiparticle, respectively.

Since the system must be charge-neutral overall, the interaction term, \(V(|\Phi|^2)\), must depend only on density fluctuations around its homogeneous value, \(\frac{\nu e B}{2\pi\hbar}\), also known as jellium model. For short-range interactions, we can approximate the interaction to be local, expressed as \(V(|\Phi|^2) = g \,(|\Phi|^2 - \frac{\nu e B}{2\pi\hbar})^2\). This potential spontaneously breaks the symmetry, leading to a vacuum expectation value for the bosonic field \(\Phi\). The dimensionless coefficient \(g / (\ell_B^2\hbar\omega_B)\) represents the ratio of the characteristic interaction energy to the bulk magnetoplasmon gap. Here,  $\ell_B=\sqrt{\hbar/(eB)}$ is the magnetic length and $\omega_B=eB/m$ is the cyclotron frequency. To avoid excitations above the gap in the magnetoplasmon spectrum, we consider, \(g / (\ell_B^2\hbar\omega_B) \ll 1\).

It is convenient to rewrite this action in terms of Madelung variables, i.e., \(\Phi = \sqrt{n}e^{i\theta}\). For simplicity, we assume a constant and uniform background magnetic field and no external electric field. Therefore, without loss of generality, we set $A_0 = 0$. The action then becomes:
\begin{align}
    S_{\text{CSGL}} = -\int & \left[\hbar n \left(\partial_t \theta + a_0\right) + \frac{\hbar^2}{2m} n \left(\partial_i \theta + a_i + \frac{e}{\hbar}A_i\right)^2 + \frac{\hbar^2}{8mn} (\partial_i n)^2 + g \left(n - \frac{\nu eB}{2\pi\hbar}\right)^2 \right. \nonumber 
    \\
    & \left. - \mu n + \frac{\hbar \nu}{4\pi} \epsilon_{ij} \left(a_0 \partial_i a_j - a_i \partial_t a_j + a_i \partial_j a_0\right)\right] d^3x\,. \label{eq:csglaction}
\end{align}

The equations of motion, obtained from the bulk variations of fields, are given by
\begin{align}
\partial_t\theta+a_0+\frac{\hbar}{2m}\left(\p_i\theta+a_i+\frac{e}{\hbar}A_i\right)^2+\frac{2g}{\hbar}\left(n-\frac{\nu e B}{2\pi\hbar}\right)-\frac{\hbar}{2m\sqrt{n}}\Delta \sqrt{n}-\frac{\mu}{\hbar} &= 0\,, \label{eq:theta}
\\
\partial_t n+\frac{\hbar}{m}\partial_i\left[n\left(\partial_i\theta+a_i+\frac{e}{\hbar}A_i\right)\right] &= 0\,, \label{eq:n}
\\
n+\frac{\nu}{2\pi}\epsilon_{ij}\partial_ia_j &= 0\,, \label{eq:Gauss}
\\
\frac{\hbar}{ m}n\left(\partial_i\theta+a_i+\frac{e}{\hbar}A_i\right) +\frac{\nu}{2\pi}\epsilon_{ij}\left(\partial_t a_j-\partial_ja_0\right) &= 0\,. \label{eq:a}
\end{align}

To reformulate these equations in hydrodynamic form, it is necessary to introduce a velocity field. A natural choice is 
\begin{align}
u_i = \frac{\hbar}{m}\left(\partial_i\theta + a_i + \frac{e}{\hbar}A_i\right), 
\label{eq:u}
\end{align}
as it maintains the continuity equation in the desired form. This is the approach is used in Ref.~\cite{stone1990superfluid}. However, we will show in the following that a choice for the velocity field which offers certain advantages in determining the hydrodynamic boundary conditions while preserving the bulk dynamics.

\subsection{Bulk Hydrodynamic Equations}

With the choice of velocity from Eq.~(\ref{eq:u}), the continuity equation becomes:
\begin{equation}
    \partial_t n + \partial_i(n u_i) = 0\,, \label{continuity-Stone}
\end{equation}
while Eq.~(\ref{eq:Gauss}) transforms into the Hall constraint:
\begin{equation}
  \epsilon_{ij}\partial_i u_j + \frac{2\pi\hbar}{\nu m}n - \frac{eB}{m} = 0\,. \label{Hall-Stone}
\end{equation}
This acts as an additional constitutive relation, linking fluid vorticity to density fluctuations relative to a constant background. The Euler equation is obtained by substituting Eq.~(\ref{eq:theta}) into Eq.~(\ref{eq:a}) and using the definition of \(u_i\). This results in
\begin{align}
    &\p_t u_{i}-\partial_i\left[\frac{u_j^2}{2}+\frac{2g}{m}\left(n-\frac{\nu eB}{2\pi}\right)-\frac{\mu}{m}-\frac{\hbar^2}{2m^2\sqrt{n}}\Delta \sqrt{n}\right]-\frac{2\pi\hbar}{\nu m}\epsilon_{ij}\,nu_j=0\,, \nonumber
    \\
    &\p_t u_i+u_j\p_ju_{i}+\frac{eB}{m}\epsilon_{ij}u_j+\p_i\left(\frac{2g}{m}n-\frac{\hbar^2}{2m^2\sqrt{n}}\Delta \sqrt{n}\right)=0\,, \label{Euler-Stone}
\end{align}
where, in the last line, we applied Eq.~(\ref{Hall-Stone}) and set the chemical potential $\mu$ to be a constant. By neglecting quantum pressure, i.e., the last term in Eq.~(\ref{Euler-Stone}), Eqs.~(\ref{continuity-Stone}-\ref{Euler-Stone}) reduce to the fluid dynamics equations for an inviscid, compressible, charged fluid in a magnetic field. They also include an additional constitutive relation that links the fluid's vorticity to density fluctuations, which explains why the composite boson condensate is considered a superfluid. These simplified equations formed the foundation of Ref.~\cite{nagaosa1994chern}. However, the complete Eq.~(\ref{Euler-Stone}), being a third-order derivative equation, requires an additional boundary condition in the presence of a hard wall, a detail that was overlooked in their analysis.

Before exploring potential hard-wall boundary conditions for Eqs.~(\ref{continuity-Stone}-\ref{Euler-Stone}), it is worth noting that these equations exhibit an unusual gradient structure in the fields. While they are first-order differential equations in the velocity fields, they are third-order in the density field. However, this system of equations can be converted into second-order differential equations by redefining the velocity field. This transformation is expressed as follows:
\begin{equation}
u_i = v_i - \frac{\hbar}{2mn} \epsilon_{ij} \partial_j n.
\end{equation}
Under this redefinition Eqs.~(\ref{continuity-Stone}) and (\ref{Hall-Stone}) become
\begin{align}
    \p_tn+\p_i(nv_i)&=0\,, \la{eq:continuity}
    \\
    \epsilon_{ij}\p_iv_j-\frac{eB}{m}+\frac{2\pi\hbar}{\nu m}n+\frac{\hbar}{2m}\p_i\left(\frac{\p_in}{n}\right)&=0\,. \la{eq:Hall}
\end{align}
After some algebra, we find that the new Euler equation can be written as:
\begin{align}
    \p_tv_i+v_j\p_jv_i=&-\frac{eB}{m}\epsilon_{ij} v_j-\frac{2g}{m}\p_in+\frac{\hbar}{2mn}\p_j\big[n\left(\epsilon_{ik}\p_k v_j+\epsilon_{jk}\p_iv_k\right)\big]\nonumber
    \\
    &-\frac{\hbar}{2m n}\p_i\left[n\frac{eB}{m}-n\,\epsilon_{jk}\p_jv_k-n\frac{\hbar}{2m}\p_j\left(\frac{\p_jn}{n}\right)\right].
\end{align}

Using Eq.~(\ref{eq:Hall}), we see that this equation becomes a second-order differential equation:
\begin{align}
    &\p_tv_i+v_j\p_jv_i+\frac{2}{m}\left(g+\frac{\pi\hbar^2}{\nu m}\right)\p_in-\frac{\hbar}{2mn}\p_j\big[n\left(\epsilon_{ik}\p_k v_j+\epsilon_{jk}\p_iv_k\right)\big]+\frac{eB}{m}\epsilon_{ij} v_j=0\,. \la{eq:Euler-1st}
\end{align}

It is often useful to introduce the fluid stress tensor \(T_{ij}\):
\begin{align}
T_{ij} = & \left[n \, \tilde{V}'(n) - \tilde{V}(n)\right] \delta_{ij} - \frac{\hbar n}{2} \left(\epsilon_{ik} \partial_k v_j + \epsilon_{jk} \partial_i v_k \right)\,. \label{eq:stress}
\end{align}
The first term in Eq.~(\ref{eq:stress}) represents the modified fluid pressure, where 
\begin{equation}
 \tilde{V}(n) = \left(g + \frac{\pi \hbar^2}{\nu m}\right) \left(n - \frac{\nu e B}{2\pi \hbar}\right)^2 
\end{equation}
is the fluid's internal energy. The second term accounts for the odd viscosity stress~\cite{avron1998odd}, with the odd viscosity coefficient being proportional to the condensate density. This stress tensor also appears in Ref.~\cite{Geracie2015-Thermal} as a result of the lowest Landau level limit \((m \rightarrow 0)\) for electrons in a magnetic field. Using the fluid stress tensor, we can rewrite the Euler equation as follows:
\be
\p_t v_i+v_j\p_j v_i=\frac{1}{m n}\p_jT_{ji}-\frac{eB}{m}\epsilon_{ij} v_j\,.  \la{eq:Euler}
\ee

The set of equations~(\ref{eq:continuity}, \ref{eq:Hall}, \ref{eq:Euler-1st}) is equivalent to Eqs.~(\ref{continuity-Stone}-\ref{Euler-Stone}). However, the former set has the advantage of comprising only second-order differential equations for both density and velocity fields. Fluid dynamic equations where the stress tensor includes at most first-order spatial derivatives are commonly referred to as first-order hydrodynamics. Additionally, as we will discuss next, boundary conditions at a hard wall are more naturally expressed in terms of the velocity field \(v_i\). Therefore, we will use the fluid dynamic equations (\ref{eq:continuity}-\ref{eq:Euler-1st}) as the basis for studying the hydrodynamics of Laughlin state and as the starting point for our analysis in this work.

\subsection{Hard-Wall Boundary Conditions}\label{sec:bc}

Any consistent set of boundary conditions for the superfluid hydrodynamic equations (\ref{eq:continuity}, \ref{eq:Hall}, \ref{eq:Euler-1st}) must respect two primary conservation laws: the conservation of the number of electrons and the conservation of the total energy of the system. Assuming 
 the Hall fluid is confined within a finite, rigid domain denoted by \( \mathcal{M} \). The number of electrons inside \( \mathcal{M} \) is determined by integrating the condensate density over the entire domain. Given that the number of electrons remains constant, we have
\begin{equation}
\frac{d}{dt} \int_{\mathcal{M}} n \, d^2x = - \oint_{\partial \mathcal{M}} n \, v_n \, ds = 0\,.
\end{equation}
This condition is satisfied if the integrand vanishes~\footnote{This is a sufficient, but not a necessary condition.}, which implies
\begin{equation}
(n v_n) \Big|_{\partial \mathcal{M}} = 0\,.
\end{equation}
Assuming the density does not vanish at the boundary, the normal component of the velocity field must be zero at the boundary. This condition, known as the no-penetration condition, ensures that no particles flow toward the hard wall.

The fluid energy is governed by an additional conservation law derived from Eqs.~(\ref{eq:continuity}, \ref{eq:Hall}, \ref{eq:Euler-1st}). This law can be expressed as:
\begin{equation}
\partial_t \mathcal{H} + \partial_i (\mathcal{H} v^i + T^{ij} v_j) = 0\,,
\end{equation}
where the energy density \(\mathcal{H}\) is defined as
\begin{equation}
\mathcal{H} = \frac{m}{2} n v_i^2 + \tilde{V}(n)\,. \la{eq:energy-dens}
\end{equation}

Therefore, the total energy of the fluid is given by the integral of \(\mathcal{H}\) over the entire domain. For simplicity, let us assume that \(\mathcal{M}\) corresponds to the lower half-plane, i.e., \(y \leq 0\). In this scenario, we have

\be
\frac{d}{dt}\int  \mathcal H\, d^2x = -\int \left[\left(\mathcal H- T_{yy}\right) v_y- T_{yx} v_x\right]\Big|_{y=0} dx. \la{dH/dt}
\ee
Energy conservation is maintained when the no-penetration condition, \(v_y(t,x,0) = 0\), is combined with either the no-stress condition, \(T_{yx}(t,x,0) = 0\), or the no-slip condition, \(v_x(t,x,0) = 0\)~\footnote{These are sufficient conditions. The necessary conditions are beyond the scope of this work and will be addressed in future studies.}. Although the composite boson model of the Laughlin state can, in principle, be supplemented with various boundary conditions, our previous work~\cite{monteiro2022topological} has shown that only the no-stress condition is compatible with the expected physics at the edge of the Laughlin state. The no-stress condition, along with the bulk continuity equation and no-penetration condition, can be reformulated as an additional dynamical equation at the edge:
\begin{align}
    T_{yx}\Big|_{y=0} = -\frac{\hbar n}{2}\left(\partial_y v_y - \partial_x v_x\right)\Big|_{y=0} = \hbar \sqrt{n}\left[\partial_t \sqrt{n} + \partial_x\left(\sqrt{n} v_x\right)\right]\Big|_{y=0} = 0\,, \la{eq:no-stress-dyn}
\end{align}
This equation can be interpreted as an additional conservation law at the edge of the sample. In this context, \(\sqrt{n}|_{y=0}\) represents the edge density, and \((\sqrt{n} v_x)|_{y=0}\) corresponds to the edge current. In our earlier work~\cite{monteiro2022topological}, we introduced a variational approach to these boundary conditions, demonstrating that the `dynamical' no-stress condition arises from a chiral boson action for an auxiliary edge field. For a comprehensive review of this variational method for deriving the hydrodynamic boundary conditions, see appendix.

\section{Weakly non-linear dynamics and method of multiple scales}
\label{sec:edgemodes}

In this section, we begin with the first-order hydrodynamic equations and derive the weakly nonlinear dynamics of the hydrodynamic edge modes. By rewriting the hydrodynamic equations in terms of the cyclotron frequency $\omega_B = eB/m$ and the magnetic length $\ell_B = \sqrt{\hbar/eB}$, we obtain the following:
\begin{align}
\p_tn + \p_i(n v^i) &= 0\,, \label{continuity} \\
\p_t v_i + v^j \p_j v_i + \epsilon_{ij} \omega_B v_j - \frac{1}{mn} \p_j T_{ji} &= 0\,,  \label{Euler} \\
n - \frac{\nu}{2\pi \ell_B^2} + \frac{\nu}{4\pi} \p_i \left( \frac{\p_i n}{n} \right) + \frac{\nu}{2\pi \ell_B^2 \omega_B} \epsilon_{ij} \p_i v_j &= 0\,, \label{Hall-constraint}
\end{align}
where the stress tensor is defined in Eq.~(\ref{eq:stress}). These bulk equations are supplemented by boundary conditions enforcing no-penetration and no-stress:
\begin{align}
v_y \Big|_{y=0} &= 0\,, \label{no-penetration} \\
\left[ \p_t \sqrt{n} + \p_x \left( \sqrt{n} v_x \right) \right] \Big|_{y=0} &= 0\,. \label{no-stress}
\end{align}

Note that the system has a characteristic length scale, $\ell_B$, and a characteristic time scale, $\omega_B^{-1}$. We are particularly interested in the surface modes described by Eqs.~(\ref{continuity}-\ref{no-stress}). Since their energy lies within the bulk gap, we focus on the long-wavelength and low-frequency regime. To help this analysis, we define the following dimensionless coordinates:
\begin{align}
\tau := \alpha \omega_B t\,, \quad
\sigma := \beta \left( \frac{x}{\ell_B} - U \omega_B t \right), \quad \xi := \frac{y}{\ell_B}\,,
\end{align}
where $\alpha$ and $\beta$ are small parameters ($\alpha, \beta \ll 1$), and $U$ is a free boost parameter. Additionally, we assume that
\begin{equation}
n(t,x,y) = n(\tau,\sigma,\xi)\,, \qquad v_i(t,x,y) = v_i(\tau,\sigma,\xi)\,.
\end{equation}

This scheme examines fluctuations in the order of the magnetic length in the $y$-direction and variations on scales much larger than the magnetic length in the $x$-direction. Let us now focus on the hydrodynamic variables and define their dimensionless counterparts:
\begin{align}
n := \frac{\nu}{2\pi \ell_B^2} \left(1 + \epsilon \, \rho \right), \quad 
v_x := \gamma \, \ell_B \omega_B \, u\,, \quad
v_y := \delta \, \ell_B \omega_B \, v\,,
\end{align}
where $\epsilon, \gamma, \delta \ll 1$. We can explore different physical regimes that emerge on different time scales depending on how these small dimensionless parameters scale relative to each other. We substitute this scaling scheme into the hydrodynamic equations to establish a consistent hierarchy of these scales, corresponding to different physical behaviors. As we will see, the Hall constraint~(\ref{Hall-constraint}) alone will determine many of the scaling relations:
\begin{align}
    \epsilon \rho+\frac{1}{2}\left(\beta^2\p_\sigma^2+\p_\xi^2\right)\ln\left(1+\epsilon\rho\right)+\beta\delta\p_\sigma v-\gamma\p_\xi u&=0\,,\nonumber
    \\
    \epsilon\left(1+\frac{1}{2}\p_\xi^2\right)\rho-\gamma\p_\xi u+\beta\left(\frac{1}{2}\beta\epsilon\p_\sigma^2\rho+\delta\p_\sigma v\right)-\frac{\epsilon^2}{4}\p_\xi^2(\rho^2)+\mathcal O(\beta^2\epsilon^2,\epsilon^3)&=0\,. \la{eq:Hall-epsilon}
\end{align}

When $\gamma$ is the largest scale, the equation enforces that $u$ must be uniform along the $\xi$ direction. Conversely, when $\epsilon$ is the dominant scale, the density fluctuation $\rho$ oscillates without decaying in the $\xi$ direction. Since we are interested in exponentially localized solutions near the edge, we set $\gamma = \epsilon$ in Eq.~(\ref{eq:Hall-epsilon}). We find that $\delta = \beta\epsilon$ by grouping the remaining linear order terms.

Next, we compare the scaling parameters $\beta^2$ and $\epsilon$. When $\epsilon \ll \beta^2$, non-linearities dominate over dispersive terms, often resulting in instabilities. However, these instabilities are counteracted by the dispersive effects. Therefore, we focus on the scaling regime where dispersive and nonlinear terms are of the same order, i.e., $\beta = \sqrt{\epsilon}$. With these relations between the scaling parameters, the Hall constraint simplifies to:
\begin{equation}
\left(1 + \frac{1}{2} \p_\xi^2\right) \rho - \p_\xi u + \epsilon \left[\frac{1}{2} \p_\sigma^2 \rho + \p_\sigma v - \frac{1}{4} \p_\xi^2 (\rho^2)\right] + \mathcal O(\epsilon^2) = 0\,. \label{eq:Hall2}
\end{equation}

The Hall constraint alone is sufficient to fix nearly all the scaling parameters, except for $\alpha$. To determine $\alpha$, we now consider the continuity equation~(\ref{continuity}) within this scaling scheme, expressed in terms of $\epsilon$:
\begin{equation}
    \epsilon \left( \alpha \p_\tau - \sqrt{\epsilon} \, U \p_\sigma \right) \rho + \epsilon \sqrt{\epsilon} \left( \p_\sigma u + \p_\xi v \right) + \epsilon^2 \sqrt{\epsilon} \left[ \p_\sigma (\rho u) + \p_\xi (\rho v) \right] = 0\,. \label{eq:rho}
\end{equation}
Note that in this equation, $\alpha$ cannot be of the same order as $\sqrt{\epsilon}$, as this has already been accounted for with the introduction of the boost parameter $U$. If $\alpha \gg \sqrt{\epsilon}$, the only time dependence of the fields would come from the boosted variable $\sigma$. To allow for more general temporal dynamics at the next-to-leading order, we set $\alpha = \epsilon^{3/2}$. This gives us our second dimensionless equation:
\begin{equation}
    \left( \p_\sigma u + \p_\xi v - U \p_\sigma \rho \right) + \epsilon \left[ \p_\tau \rho + \p_\sigma (\rho u) + \p_\xi (\rho v) \right] = 0\,. \label{eq:rho2}
\end{equation}

All scaling parameters are thus fixed in terms of \(\epsilon\). We summarize the scales and their corresponding variables in Table \ref{tab:scaling}.
\begin{table}[]
    \centering
    \begin{tabular}{|c|c|c|}
        \hline
    Scale   &  Dependence on $\epsilon$ & Variable  \\
    \hline
    $\epsilon$   & $\epsilon$  & $n=\frac{\nu}{2\pi \ell_B^2}\left(1+\epsilon \,\rho\right)$\\
     \hline
      $\alpha$   &  $\epsilon^{3/2}$ &   $\tau=\epsilon^{3/2} \omega_B t$\\
       \hline
      $\beta$   & $\epsilon^{1/2}$  & $\sigma=\epsilon^{1/2}(x/l_B-U \omega_B t)$\\
       \hline
      $\gamma$   & $\epsilon$  & $v_x=\epsilon l_B \omega_B u$\\
       \hline
        $\delta$   & $\epsilon^{3/2}$  & $v_y=\epsilon^{3/2} l_B \omega_B v$\\
        \hline
    \end{tabular}
    \caption{Counting scheme adopted for the fields and space-time variables. This scaling assumption resembles the one used to derive the KdV equation for shallow water dynamics.}
    \label{tab:scaling}
\end{table}

Applying this scaling assumption to the two components of the Euler equation yields:
\begin{align}
    & \left[c^2 \p_\sigma \rho - U \p_\sigma u + \left(1 + \frac{1}{2} \p_\xi^2 \right) v\right] \nonumber \\
    & \quad + \epsilon \left[\p_\tau u + u \p_\sigma u + v \p_\xi u + \frac{1}{2} \left(\p_\xi v - \p_\sigma u \right) \p_\xi \rho + \frac{1}{2} \p_\xi u \p_\sigma \rho + \frac{1}{2} \p_\sigma^2 v \right] + \mathcal O(\epsilon^2) = 0\,, \la{eq:u}
\end{align}
for the $x$-component and
\begin{align}
    & c^2 \p_\xi \rho - \left(1 + \frac{1}{2} \p_\xi^2 \right) u - \epsilon \left(\frac{1}{2} \p_\sigma^2 u + U \p_\sigma v + \frac{1}{2} \p_\xi u \p_\xi \rho \right) + \mathcal O(\epsilon^2) = 0\,. \la{eq:v}
\end{align}
for the $y$-component. Note that under this particular scaling, Eq.~(\ref{eq:v}) is no longer a dynamical equation.

The dimensionless speed of sound \(c\), introduced in the previous equations, is defined in terms of the ratio between the characteristic interaction energy and the cyclotron gap in the magnetoplasmon dispersion. Specifically, it is given by \(c^2 = 1 + \frac{\nu \lambda}{\pi \ell_B^2 \hbar \omega_B}\). To avoid excitations above the cyclotron gap, we require \(\frac{\nu \lambda}{\pi \ell_B^2 \hbar \omega_B} < 1\), which implies that \(1 < c^2 < 2\).

The no-penetration condition~(\ref{no-penetration}) is straightforward and gives us
\be
v\Big|_{\xi=0}=0\,, \la{eq:v-edge}
\ee
while the no-stress condition~(\ref{no-stress}) becomes
\begin{align}
     \left[\p_\sigma u - \frac{U}{2}\p_\sigma\rho + \frac{\epsilon}{2}\left(\p_\tau\rho + \p_\sigma(\rho u) + \frac{c}{4}\partial_\sigma(\rho^2)\right) + O(\epsilon^2)\right]\bigg|_{\xi=0}=0\,. \la{eq:rho-edge}
\end{align}
The no-stress boundary condition is a dynamical equation defined at the hard wall edge, i.e., at $\xi=0$. This edge equation involves the fields $u$ and $\rho$. To fully describe the edge dynamics solely in terms of the density $\rho$, we must express the $x$-component of the velocity field, $u$, in terms of the density $\rho$ and its derivatives by solving the bulk equations.

\begin{figure*}
\centering
\includegraphics[scale=0.35]{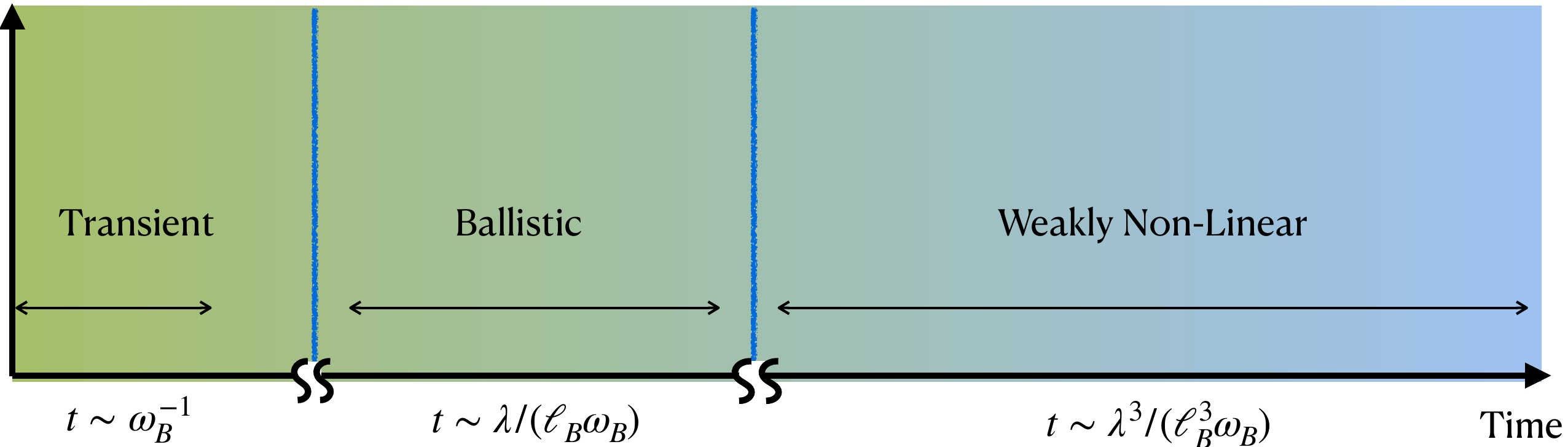}
\caption{Schematic of time scales.}
\label{fig:schematic}
\end{figure*}

\subsection{Ballistic Regime} \la{sec:linear}

The variable \(\sigma = \sqrt{\epsilon} (x/\ell_B - U \omega_B t)\) represents the boosted coordinate, where \(U\) is the group velocity of the ballistic wave. Dispersive and nonlinear terms evolve on the slower timescale \(\tau = \sqrt{\epsilon^3}\omega_B t\). This means ballistic effects dominate the system's early-time dynamics, specifically for \(t \sim 1/(\sqrt{\epsilon}\omega_B)\), while dispersive and nonlinear effects emerge at later times, around \(t \sim 1/(\sqrt{\epsilon^3}\omega_B)\).

The parameter \(\sqrt{\epsilon}\) describes the ratio between the magnetic length and the characteristic wavelength \(\lambda\) of the system, as can be inferred from the definition of \(\sigma\). We can organize the dynamics of Laughlin edge states in terms of three distinct regimes: a fast, transient regime \((t \sim 1/\omega_B)\), where excitations above the gap are important\footnote{This window shrinks as the external magnetic field increases.}; a ballistic regime observed at earlier times \((t \sim \lambda/(\ell_B \omega_B))\); and a nonlinear regime at much later times \((t \sim \lambda^3/(\ell_B^3 \omega_B))\), as schematically depicted in Fig.~\ref{fig:schematic}.

The presence of a gap introduces a high frequency and a small length scale to the system. Therefore, as long as we probe small frequencies and long wavelengths, physical quantities can be expressed as derivative expansions. For example, this allows us to expand the velocity fields as a series in powers of \(\epsilon\), with the coefficients being local functions of the density fluctuation \(\rho\).

Before examining the weakly nonlinear regime, let us address the leading-order solutions. This corresponds to the ballistic regime, where we truncate the equations at \(\mathcal{O}(1)\). In this leading order, the non-dynamical bulk equations become:
\begin{align}
    & \left(\p_\xi^2+2\right)\rho-2\p_\xi u+\mathcal{O}(\epsilon)=0\,, \la{eq-Hall-linear}
    \\
     &\left(\p_\xi^2+2\right)u-2c^2\p_\xi\rho+\mathcal{O}(\epsilon)=0\,, \la{eq-v-linear}
\end{align}
while the dynamical bulk equations are:
\begin{align}
  &\left(\p_\xi^2+2\right)v+2c^2\p_\sigma\rho-2U\p_\sigma u+\mathcal{O}(\epsilon)=0\,, \la{eq-u-linear}
    \\
    &  \p_\xi v+\p_\sigma u-U\p_\sigma\rho+\mathcal{O}(\epsilon)=0\,. \la{eq-rho-linear}
\end{align}

The strategy is to use the leading-order equations to express \( u \) and \( v \) in terms of \( \rho \) up to order \( \epsilon \). The leading-order equations do not capture any dynamics but serve to determine the boundary layer structure of the system. For example, by combining Eqs.~(\ref{eq-Hall-linear}) and (\ref{eq-v-linear}), we can express \( u \) in terms of \( \rho \), as follows:
\be
\left(\p_\xi^2 + 2c\,\p_\xi + 2\right)\left(u - c\,\rho\right) + \mathcal{O}(\epsilon) = 0 \,, \qquad \Rightarrow \qquad u = c\,\rho + \mathcal{O}(\epsilon) \,. \la{u0-solution}
\ee
Here, we utilize the fact that the homogeneous solutions of the operator \( (\p_\xi^2 + 2c\,\p_\xi + 2) \) are unbounded as \( \xi \to -\infty \). Substituting this new constitutive relation~(\ref{u0-solution}) into Eqs.~(\ref{eq-v-linear}) and (\ref{eq-rho-linear}), we obtain:
\begin{align}
    & v = \left(U - c\right)\left(c - \frac{1}{2}\p_\xi\right)\p_\sigma\rho + \mathcal{O}(\epsilon)\,, \la{v0-solution}
    \\
    & \left(\p_\xi^2 - 2c\,\p_\xi + 2\right)\rho + \mathcal{O}(\epsilon) = 0\,. \la{rho0-solution}
\end{align}

\subsection{Boundary Conditions and Two Edge Modes}
Notice that Eq.~(\ref{rho0-solution}) is explicitly a second-order differential equation in \( \rho \), which supports our claim that this system of bulk equations requires two boundary conditions. By truncating the boundary conditions~(\ref{eq:v-edge}) and (\ref{eq:rho-edge}) at leading order, we get:
\begin{align}
  \left(2\p_\sigma u - U \p_\sigma \rho\right)\Big|_{\xi=0} + \mathcal{O}(\epsilon) = 0 \,, & \la{eq-rho-edge-linear}
    \\
     v\Big|_{\xi=0} = 0 \,. & \la{eq-v-edge-linear}  
\end{align}

The no-stress condition~(\ref{eq-rho-edge-linear}) can be expressed solely in terms of \( \rho \) at the edge by substituting \( u \) from Eq.~(\ref{u0-solution}). This gives:
\begin{align}
     &\left(2c - U\right)\p_\sigma \rho \Big|_{\xi=0} + \mathcal{O}(\epsilon) = 0\,. \la{U-solution}
\end{align}
Notably, if \( U \neq 2c \), Eq.~(\ref{U-solution}) implies that \( \rho|_{\xi=0} \sim \mathcal{O}(\epsilon) \). Similarly, the no-penetration condition~(\ref{eq-v-edge-linear}) can be written in terms of \( \rho \) by using Eq.~(\ref{v0-solution}):
\begin{align}
&\left(U - c\right)\left(2c - \p_\xi\right)\p_\sigma \rho \Big|_{\xi=0} + \mathcal{O}(\epsilon) = 0\,. \la{eq:nopen-sol}
\end{align}
This equation implies that, for \( U \neq c \), we have \( \p_\xi \rho|_{\xi=0} = 2c \rho|_{\xi=0} + \mathcal{O}(\epsilon) \).

In the scenario when \( U \neq 2c \) and \( U \neq c \) are satisfied simultaneously, the boundary conditions impose both \( \rho|_{\xi=0} \sim \mathcal{O}(\epsilon) \) and \( \p_\xi \rho|_{\xi=0} \sim \mathcal{O}(\epsilon) \). This implies that the density fluctuation \( \rho \) itself is of order \( \epsilon \), which contradicts the initial scaling assumption. Consequently, the analysis splits into two distinct cases: either \( U = c \), referred to as the Kelvin mode (owing to its connection to the kelvin mode in shallow water equations~\cite{monteiro2023coastal}), or \( U = 2c \), known as the chiral boson mode.

Before analyzing each mode individually, it is worth noting that the chiral boson mode disappears when the no-slip condition replaces the no-stress boundary condition. In this case, Eq.~(\ref{eq-rho-edge-linear}) is replaced by
\begin{align}
    u\Big|_{\xi=0} = c\rho\Big|_{\xi=0} + \mathcal{O}(\epsilon) = 0\,.
\end{align}
This shows that the Kelvin mode persists in the system simply by imposing the no-penetration condition which is typically satisfied by setting $v=0$ everywhere. Recent works have shown that this mode appears in disparate systems, such as shallow water equations with coastal boundary conditions, magnetoplasmons~\cite{jin2016topological}, active matter~\cite{poggioli2023emergent}, and classical plasma subject to magnetic fields~\cite{souslov2019topological}. The robustness of the Kelvin mode has been studied across various systems, including magnetoplasmons with long-range Coulomb interactions, deformed boundary conditions~\cite{iga1995transition, tauber2020anomalous}, and interface Kelvin modes, where the magnetic field or the Coriolis parameter switches sign~\cite{tauber2019bulk}. Within the context of the CSGL theory, we established that the Kelvin mode remains non-dispersive to all orders in wave number and is not associated with the chiral edge dynamics expected at the edge of a Laughlin state~\cite{monteiro2023coastal}. Intuitively, in condensed matter systems with bulk-edge correspondence, edge modes must disperse at some wavenumber to merge into the bulk, which is not the case with the Kelvin mode. Furthermore, we showed that imposing any linearly independent second boundary condition eliminates fluctuations of the density and velocity fields at the edge, and the Kelvin mode governs near-edge dynamics. In Sec.~\ref{sec:Kelvin}, we will further discuss the weakly non-linear dynamics associated with the near-edge behavior of the Kelvin mode.

In contrast, the chiral boson mode emerges directly from the edge continuity equation~(\ref{no-stress}), which results from an additional \( U(1) \) symmetry at the system's edge. This symmetry is realized through the auxiliary chiral boson introduced in Sec.~\ref{sec:review}. Moreover, this \( U(1) \) symmetry becomes anomalous in the presence of a tangential electric field.

This supports our claim that the chiral boson mode is indeed the edge mode described by Wen's theory, while the Kelvin mode lacks any apparent associated \( U(1) \) symmetry. This is precisely why the Kelvin mode does not linearly couple to the tangential electric field, as shown in Ref.~\cite{monteiro2023coastal}, nor does it exhibit diffusion when dissipation is introduced at the FQH edge, as shown in Ref.~\cite{monteiro2023kardar}.

\section{Chiral Boson Mode} \la{sec:CBmode}

As shown in the previous section, the leading-order equations determine the boundary layer profile of the fields and their ballistic dynamics. For the chiral boson, in addition to \( u = c\rho + \mathcal{O}(\epsilon) \), we also find:
\begin{align}
    & \rho = \varrho(\tau, \sigma) e^{c\xi} \left[\cos\left(\sqrt{2 - c^2}\,\xi\right) + \frac{c}{\sqrt{2 - c^2}} \sin\left(\sqrt{2 - c^2}\,\xi\right)\right] + \mathcal{O}(\epsilon), \la{u0-CB}
    \\
    & v = \p_\sigma \varrho(\tau, \sigma) \frac{c\, e^{c\xi}}{\sqrt{2 - c^2}} \sin\left(\sqrt{2 - c^2}\,\xi\right) + \mathcal{O}(\epsilon), \la{v0-CB}
\end{align}
where \(\varrho = \rho|_{\xi=0}\). However, these explicit forms will not be necessary for the remainder of the analysis.

We have established that the velocity field can be expanded in powers of \(\epsilon\), with the expansion coefficients expressed as local functions of the density fluctuations. According to that, the velocity components can be written as:
\begin{align}
    u &= c\rho + \epsilon u_1 +\mathcal O(\epsilon^2)\,,  \label{eq:u-expansion} \\
    v &= c\left(c - \frac{1}{2}\p_\xi\right)\p_\sigma \rho + \epsilon v_1+\mathcal O(\epsilon^2)\,, \label{eq:v-expansion}
\end{align}
where \(u_1\) and \(v_1\) are functions of \(\rho\) and its derivatives, determined by the bulk equations. To derive the weakly nonlinear dynamics of the chiral boson, we substitute Eqs.~(\ref{eq:u-expansion}) and (\ref{eq:v-expansion}) into the first-order Eqs.~(\ref{eq:Hall2}-\ref{eq:rho-edge}), setting \(U = 2c\) to account for the mode's group velocity.

\subsection{Korteweg de-Vries Dynamics}

The no-stress boundary condition~(\ref{eq:rho-edge}) governs the dynamics of the chiral boson mode. Thus, we only need to determine \(u_1\) in terms of \(\rho\), as \(v_1\) does not contribute to the chiral boson's dynamical equation. To find \(u_1\), we focus on the non-dynamical equations:
\begin{align}
    &\left(\p_\xi^2 - c\p_\xi + 2\right)\rho + \epsilon\left[\left(2c^2 - c\p_\xi + 1\right)\p_\sigma^2\rho - \frac{1}{2}\p_\xi^2(\rho^2) - \p_\xi u_1\right] + \mathcal O(\epsilon^2) = 0\,, \la{eq:rho-u1}
    \\
    &c\left(\p_\xi^2 - c\p_\xi + 2\right)\rho + \epsilon\left[c\left(4c^2 - 2c\p_\xi + 1\right)\p_\sigma^2\rho + c(\p_\xi \rho)^2 + \left(\p_\xi^2 + 2\right)u_1\right] + \mathcal O(\epsilon^2) = 0\,.
\end{align}

By subtracting the second equation from \(c\) times the first, we obtain:
\begin{align}
    &\left(\p_\xi^2 +2c\p_\xi + 2\right)u_1 + c\left[2(\p_\xi\rho)^2 + \rho\p_\xi^2\rho\right] + c^2\left(2c - \p_\xi\right)\p_\sigma^2\rho + \mathcal{O}(\epsilon) = 0. \la{eq:u1}
\end{align}
The goal is to solve for \(u_1\) in terms of \(\rho\) up to order \(\epsilon\). Since \((\p_\xi^2 - 2c\p_\xi + 2)\rho \sim \mathcal{O}(\epsilon)\), we can write:
\begin{align}
    &\left(\p_\xi^2 + 2c\p_\xi + 2\right)\left(\rho^2 + c\rho\p_\xi\rho\right) = (4c^2 + 1)\left[2(\p_\xi\rho)^2 + \rho\p_\xi^2\rho\right] + \mathcal{O}(\epsilon)\,, \la{eq:identity}
    \\
    &\left(\p_\xi^2 + 2c\p_\xi + 2\right)(2c^2 - 1 - c\p_\xi)\p_\sigma^2\rho = 4c\left(2c - \p_\xi\right)\p_\sigma^2\rho + \mathcal{O}(\epsilon)\,.
\end{align}

Substituting these results into Eq.~(\ref{eq:u1}) and using the fact that the operator \((\p_\xi^2 + 2c\p_\xi + 2)\) does not admit bounded solutions in the lower half-plane, we arrive at:
\begin{equation}
    u_1 = \frac{c}{4}(c\p_\xi + 1 - 2c^2)\p_\sigma^2\rho - \frac{c}{4c^2 + 1}\left(\rho^2 + c\rho\p_\xi\rho\right) + \mathcal{O}(\epsilon)\,. \la{eq:u1(rho)}
\end{equation}

This expression for \(u_1\) implies that the density fluctuation \(\rho\) satisfies the following bulk equation:
\begin{align}
    (\p_\xi^2 - 2c\p_\xi + 2)\rho + \epsilon\left[\p_\sigma^2\rho + \frac{3c}{2}(2c - \p_\xi)\p_\sigma^2\rho - \frac{2c^2 + 1}{2(4c^2 + 1)}\p_\xi^2\left(\rho^2\right) + \frac{4c}{4c^2 + 1}\rho\p_\xi\rho\right] + \mathcal O(\epsilon^2) = 0\,. \la{eq:rho4}
\end{align}
Note that the bulk equation~(\ref{eq:rho4}) for \(\rho\) is not dynamical; it simply defines the density profile. The expression for \(u_1\) together with Eq.~\ref{eq:u-expansion} and $U=2c$ can then be substituted in the dynamical boundary condition given in Eq.~\ref{eq:rho-edge} to obtain the KdV equation for the edge dynamics: 
\begin{tcolorbox}[colback=green2!5!white,colframe=green2,title={\bf KdV equation}]
\begin{equation}
    \left[\p_\tau\rho + \frac{c}{2}\p_\sigma^3\rho + \frac{4c^2 - 1}{4c^2 + 1}\,c\rho\,\p_\sigma\rho + \mathcal{O}(\epsilon)\right]\bigg|_{\xi=0} = 0 \,, \la{eq:KdV} 
\end{equation}
\end{tcolorbox}
where, in the last equation, we used the fact that the chiral boson mode satisfies \(\p_\xi\rho|_{\xi=0} = 2c\rho|_{\xi=0} + \mathcal{O}(\epsilon)\). This is the central result of this paper where we showed that in the weakly non-linear limit, the edge density profile follows the Korteweg-de Vries (KdV) equation, which is integrable and supports soliton solutions.

\subsection{Hamiltonian Structure}

The KdV edge dynamics (Eq.~(\ref{eq:KdV})) derived in the previous section can be interpreted as an equation of motion for the boundary value of the density fluctuations. The effective edge Hamiltonian associated with this edge dynamics can be deduced from the energy density of this edge mode. By imposing the existence of a Hamiltonian structure in the system, we can determine the algebra between edge densities, which will enable us to make contact with the Kac-Moody algebra associated with the $\chi$LL theory.  It is natural to assume that Eq.~(\ref{eq:KdV}) retains the Hamiltonian structure of the system, where the Hamiltonian corresponds to the total energy of the fluid. For this edge mode, we expect the fluid's energy density to be exponentially localized near the boundary and to depend solely on the density fluctuation at the edge. Thus, the Poisson brackets for the edge density fluctuation can be derived by imposing the condition:
\begin{equation}
    \p_t\varrho(t, \sigma) := \{\varrho(t, \sigma), H\} = \int d\sigma'\{\varrho(t, \sigma), \varrho(t, \sigma')\}\frac{\delta H}{\delta\varrho(t, \sigma')}\,. \label{eq:EoM-PB}
\end{equation}
Here, for simplicity, we defined \(\rho|_{\xi=0} = \varrho\). Using Eq.~(\ref{eq:KdV}), the left-hand side of Eq.~(\ref{eq:EoM-PB}) becomes:
\begin{equation}
    \p_t\varrho = \sqrt{\epsilon} \omega_B\left(-2c \p_\sigma \varrho + \epsilon \p_\tau \varrho\right) = -\sqrt{\epsilon} \omega_B \left[2c \p_\sigma \varrho + \epsilon \frac{c}{2} \p_\sigma^3 \varrho + \epsilon \frac{4c^2 - 1}{4c^2 + 1} c \rho\, \p_\sigma \varrho+\mathcal O(\epsilon^2)\right], \la{eq:edge-dynamics}
\end{equation}
whereas the system's Hamiltonian is obtained by substituting the expressions from Eqs.~(\ref{eq:u-expansion}) and (\ref{eq:v-expansion}) into the energy density defined in Eq.~(\ref{eq:energy-dens}). This yields:
\begin{equation}
    \mathcal{H} = \frac{\nu \hbar \omega_B}{2\pi \ell_B^2} \epsilon^2\left[c^2 \rho^2 + \epsilon\left(\frac{c^2}{2} \rho^3 + \frac{c^2}{8} \left(2c \p_\sigma \rho - \p_\xi \p_\sigma \rho\right)^2 + c \rho u_1\right) + O(\epsilon^2)\right], \la{eq:Hdensity}
\end{equation}
where \(u_1\) is given by Eq.~(\ref{eq:u1(rho)}) and \(\rho\) satisfies Eq.~(\ref{eq:rho4}).

After some tedious algebra, we can show that by using Eq.~(\ref{eq:rho-u1}), we obtain the following expression:
\begin{align}
    c^2\rho^2 + \epsilon \rho u_1 &= \frac{c}{8} \partial_\xi\left[\left(\p_\xi\rho - 2c\rho\right)\left(\p_\xi\rho - 2c\rho - 4\epsilon u_1 - 2\epsilon\rho \p_\xi\rho\right) + 2\rho^2 - \epsilon \frac{4\rho^3}{3} + \epsilon\left(\p_\xi\p_\sigma\rho - 2c\p_\sigma\rho\right)^2 \right] \nonumber
    \\
    &\quad + \epsilon \frac{c}{4} \p_\sigma \left[\left(\p_\xi\rho - 2c\rho\right)\left((2c^2+1)\rho - c\p_\xi\rho\right)\right] + \epsilon \frac{c^2}{4}\left(2c \p_\sigma\rho - \p_\xi\p_\sigma\rho\right)^2 + \mathcal{O}(\epsilon^2)\,,
\end{align}
and also:
\begin{align}
    \frac{c}{4}\left(2c \p_\sigma\rho - \p_\xi\p_\sigma\rho\right)^2 = \p_\xi \left[(2c^2 + 1)\left(2c \p_\sigma\rho - \p_\xi\p_\sigma\rho\right)^2 - \p_\sigma\rho \left(2c \p_\sigma\rho - \p_\xi\p_\sigma\rho\right) + \frac{(\p_\sigma\rho)^2}{8}\right] + \mathcal{O}(\epsilon).
\end{align}

Substituting these identities into the fluid Hamiltonian expression, and imposing the boundary condition \(\p_\xi\rho|_{\xi=0} = 2c\varrho + \mathcal{O}(\epsilon)\), the system Hamiltonian becomes:
\begin{align}
    H &= \frac{\nu\hbar\omega_B}{8\pi} \epsilon^{3/2} \int d\sigma \left[c \varrho^2 + \epsilon \frac{3c}{4} (\p_\sigma \varrho)^2 + \epsilon \frac{2c}{3} \left(\frac{4c^2+3}{4c^2+1}\right) \varrho^3 + \mathcal{O}(\epsilon^2)\right] \,. \la{eq:Hamiltonian-edge}
\end{align}
Varying this Hamiltonian gives:
\begin{equation}
    \frac{\delta H}{\delta \varrho} = \frac{\nu\hbar\omega_B}{8\pi} \epsilon^{3/2} \left[ 2c\varrho + \epsilon\left(-\frac{3c}{2} \p_\sigma^2 \varrho + 2c \frac{4c^2+3}{4c^2+1} \varrho^2 \right)\right] + \mathcal{O}(\epsilon^{9/2})\,,
\end{equation}
Substituting this into Hamilton's equation~(\ref{eq:EoM-PB}) and comparing it with the dynamical equation~(\ref{eq:edge-dynamics}), we find that:
\begin{align}
   &-\frac{\nu\hbar}{8\pi}\epsilon \int d\sigma'\{\varrho(\sigma),\varrho(\sigma')\}\left[2c\varrho(\sigma')+\epsilon\left(-\frac{3c}{2}\p_{\sigma'}^2\varrho(\sigma')+2c\,\frac{4c^2+3}{4c^2+1}\varrho^2(\sigma')\right)\right]\nonumber
   \\
   &=2c\p_\sigma\varrho+\epsilon\frac{c}{2}\p_\sigma^3\varrho+\epsilon\frac{4c^2-1}{4c^2+1}\,c\,\varrho\,\p_\sigma\varrho+O(\epsilon^2)\,. \la{eq:ham-eq}
\end{align}
By comparing the leading-order term, we conclude that 
\begin{tcolorbox}[colback=green2!5!white,colframe=green2,title={\bf $U(1)$ Kac-Moody Algebra}]
\begin{equation}
    \{\varrho(\sigma), \varrho(\sigma')\} = -\frac{8\pi}{\nu\hbar\epsilon}\p_\sigma\delta(\sigma - \sigma') + \mathcal{O}(1)\,. \la{eq:Kac-Moody}
\end{equation}
\end{tcolorbox}
This reproduces Wen's $\chi$LL result in the leading order and it is often referred to as $U(1)$ Kac-Moody algebra~\cite{wen1990chiral}. However, this alone does not yield the correct coefficients for the dispersive and nonlinear terms. Therefore, the bracket \(\{\varrho(\sigma), \varrho(\sigma')\}\) requires \(\epsilon\) corrections and can only reproduce the equation of motion by neglecting terms of order \(\mathcal{O}(\epsilon^2)\). Thus, the antisymmetric bracket that satisfies Eq.~(\ref{eq:ham-eq}) up to order $\epsilon$ is given by:
\begin{tcolorbox}[colback=green2!5!white,colframe=green2,title={\bf $\epsilon$-Corrections to the $U(1)$ Kac-Moody Algebra}]
\begin{align}
    \{\varrho(\sigma), \varrho(\sigma')\} = -\frac{8\pi}{\nu\hbar\epsilon}\left[1 + \epsilon \p^2_\sigma - \frac{\epsilon}{6}\frac{8c^2 + 13}{4c^2 + 1} \left(\varrho(\sigma) + \varrho(\sigma')\right)\right]\p_\sigma \delta(\sigma - \sigma') + \mathcal{O}(\epsilon)\,.
\end{align}
\end{tcolorbox}
Notably, this bracket satisfies the Jacobi identity up to order $\epsilon$. However, it is inconvenient to rely on a Hamiltonian structure that holds only when the system is truncated. In the following, we will introduce a change of variables that eliminates the terms proportional to \(\mathcal O(1)\) in the Poisson algebra.

\subsection{KdV variables and Quantization}

The Hamiltonian structure discussed in the previous section closes only up to $\mathcal O(\epsilon^2)$. However, to quantize this system, it is convenient to introduce new variables in which the Hamiltonian structure closes without truncation. To achieve this, the new variables \(\mathfrak{H}_\epsilon\) must satisfy the following relation:
\begin{equation}
    \{\mathfrak H_{\epsilon}(\sigma),\mathfrak H_{\epsilon}(\sigma')\}=-\frac{2\pi}{\hbar\nu\ell_B^2\epsilon}\partial_\sigma\delta(\sigma-\sigma')+\mathcal{O}(\epsilon)\,. \la{eq:PB-KdV}
\end{equation}
The KdV variable \(\mathfrak{H}_{\epsilon}\) can be expressed in terms of \(\varrho\) as follows:
\begin{equation}
     \mathfrak H_{\epsilon}=\frac{1}{2\ell_B}\left(\varrho-\frac{\epsilon}{2}\p_\sigma^2\varrho+\frac{\epsilon}{12}\frac{8c^2+13}{4c^2+1}\varrho^2\right)+\mathcal O(\epsilon^2)\,.
\end{equation}
And the Hamiltonian, in terms of \(\mathfrak{H}_{\epsilon}\), takes the well-known KdV form:
\begin{align}
        H&=\frac{\nu\hbar\omega_B\ell_B^2c}{2\pi}\epsilon^{3/2}\int d\sigma\left[\mathfrak H_\epsilon^2-\epsilon\,\frac{\ell_B}{4}(\p_\sigma\mathfrak H_\epsilon)^2+\epsilon\,\frac{\ell_B}{3}\left(\frac{4c^2-1}{4c^2+1}\right)\mathfrak H_\epsilon^3+\mathcal O(\epsilon^{2})\right]\,, \la{eq:H-KdV}
\end{align}

The Hamiltonian structure closes without relying on truncation when expressed in terms of the variable $\mathfrak H_\epsilon$. To illustrate this, we can write the equation of motion for \(\mathfrak{H}_{\epsilon}\) as follows:
\begin{equation}
     \p_t\mathfrak H_\epsilon=\sqrt{\epsilon}\omega_B(-2c\p_\sigma\mathfrak H_\epsilon+\epsilon \p_\tau\mathfrak H_\epsilon)=-\sqrt{\epsilon}\omega_Bc\left[2\p_\sigma\mathfrak H_\epsilon+\frac{\epsilon}{2}\p_\sigma^3\mathfrak H_\epsilon+\epsilon\frac{4c^2-1}{4c^2+1}\,\mathfrak H_\epsilon\,\p_\sigma\mathfrak H_\epsilon+\mathcal O(\epsilon^{2})\right].
\end{equation}
The right-hand side of this equation precisely corresponds to \(\{\mathfrak{H}_{\epsilon}, H\}\) without relying on any truncation. Therefore, we can neglect higher-order terms in the Hamiltonian \((\ref{eq:H-KdV})\) as well as in the Poisson bracket \((\ref{eq:PB-KdV})\), since they do not contribute to the equations of motion. Expressing the bracket as a function of the coordinate \(x\), we obtain:
\begin{equation}
    \{\mathfrak H_\epsilon(x),\mathfrak H_\epsilon(x')\}=-\frac{2\pi}{\hbar\nu\epsilon^2}\partial_x\delta(x-x')\,.
\end{equation}

Note that we can eliminate the $\epsilon$ dependence in the bracket by defining $\mathfrak H=\epsilon\mathfrak H_\epsilon$. Furthermore, this Poisson bracket is nearly canonical, allowing us to quantize the system by applying the Dirac prescription. This procedure promotes the bracket to the following commutator:
\begin{equation}
    [\mathfrak H(x),\mathfrak H(x')]=-\frac{2\pi i}{\nu}\partial_x\delta(x-x')\,.
\end{equation}
Consequently, the system Hamiltonian takes the form:
\begin{align}
        H&=\frac{\nu\hbar \omega_B\ell_B c}{2\pi}\int dx\left[\mathfrak H^2-\frac{\ell_B^2}{4}(\p_x\mathfrak H)^2+\frac{\ell_B}{3}\left(\frac{4c^2-1}{4c^2+1}\right)\mathfrak H^3\right]\,,
\end{align}
where the terms \(\mathfrak{H}^2\) and \(\mathfrak{H}^3\) should be defined using the appropriate regularization scheme.

\section{Kelvin Mode} \la{sec:Kelvin}

In our recent work~\cite{monteiro2023coastal, monteiro2023kardar}, we showed that the edge dynamics of the CSGL theory are described by the chiral boson mode. While this mode accounts for the (gauge) anomalous transport at the sample edge and diffuses in the presence of dissipation at the boundary, the Kelvin mode remains inert in the linearized regime. Yet, the Kelvin mode is also a solution to the boundary value problem of the composite boson model, and for completeness, we devote this section to analyzing the Kelvin mode's weakly nonlinear dynamics within the CSGL theory. In the context of geophysics, the weakly non-linear dynamics of the equatorial Kelvin modes and the related instabilities were extensively studied by Boyd~\cite{boyd1978effects1, boyd1978effects2}. Our previous works have revealed the similarities and differences between the Kelvin mode emerging in the shallow water equations and the CSGL theory with a hard wall in Ref.~\cite{monteiro2023coastal}. 

At leading order, the Kelvin mode is characterized by $U=c$, $v\sim\mathcal O(\epsilon)$ and $\rho |_{\xi=0}\sim \mathcal O(\epsilon)$. Furthermore, the density fluctuation takes the form:
\begin{align}
    & \rho=\eta(\tau,\sigma)\, e^{c\xi}\sin\left(\sqrt{2-c^2}\,\xi\right)+\mathcal{O}(\epsilon). \la{rho0-K}
\end{align}
Here, we identify the function $\eta(\tau,\sigma)$ with $\p_\xi\rho\big|_{\xi=0}$ as
\begin{align}
\p_\xi\rho\big|_{\xi=0}=\sqrt{2-c^2}\,\eta(\tau,\sigma)+\mathcal O(\epsilon)\,.
\end{align}

Following the same steps from the last section, the velocity components $u$ and $v$ can be written in terms of density fluctuations $\rho$ as:
 \begin{align}
    u &= c\rho + \epsilon u_1 +\mathcal{O}(\epsilon^2)\,,  \label{eq:u-expansion-kelvin} \\
    v &=0 + \epsilon v_1+\mathcal O(\epsilon^2)\,, \label{eq:v-expansion-kelvin}
\end{align}
where \(u_1\) and \(v_1\) are functions of \(\rho\) and its derivatives, determined by the bulk equations. To determine 
\(u_1\) for the Kelvin mode, we substitute Eqs.~(\ref{eq:u-expansion-kelvin}) and (\ref{eq:v-expansion-kelvin}) into the non-dynamical  Eqs.~(\ref{eq:Hall2}-\ref{eq:rho-edge}) with \(U = c\) and obtain:
\begin{align}
&\left(\p_\xi^2 - 2c\p_\xi + 2 \right) \rho  + \epsilon \left[ \p_\sigma^2 \rho -\p_\xi u_1 - \frac{1}{2} \p_\xi^2 (\rho^2)\right] + \mathcal O(\epsilon^2) = 0,\la{eq:hall-kelvin}\\
& c\left(\p_\xi^2 - 2c\p_\xi + 2 \right) \rho  + \epsilon \left[ \p_\sigma^2 \rho +\left( \p_\xi^2+2\right)u_1+  (\p_\xi \rho)^2 \right] + \mathcal{O}(\epsilon^2) = 0\,. \la{eq:v}
\end{align}
The procedure here is very similar to the one from the last section. Hence, using Eq.~(\ref{eq:identity}), we get
\begin{equation}
 u_1=-\frac{c}{4c^2+1}\left(\rho^2+c\rho\,\p_\xi\rho\right)+\mathcal{O}(\epsilon)\,. \la{eq:u1(rho)}
\end{equation}

Let us now turn our attention to the dynamical equations~(\ref{eq:rho2}) and (\ref{eq:u}). Substituting the expressions for $u$ and $v$:
\begin{align}
     \p_\xi v_1+\p_\sigma u_1+\p_\tau\rho+c\rho\,\p_\sigma\rho+\mathcal{O}(\epsilon)&=0 \label{eq:rho3}\,,
     \\
 \left(\p_\xi^2+2\right)v_1-2c\p_\sigma u_1+2c\,\p_\tau \rho+2c^2\rho\p_\sigma \rho+\mathcal{O}(\epsilon^2)&=0\,.  \label{eq:u2}  
\end{align}
We can solve $v_1$ by differentiating Eq.~(\ref{eq:u2}) in terms of $\xi$ and subtracting it by Eq.~(\ref{eq:rho3}). This yields:
\begin{align}
 v_1=\left[\left(\frac{1}{2}\p_\xi-c\right)\left(\p_\tau\rho+c\rho\,\p_\sigma\rho\right)-\frac{c}{4c^2+1}\left(\frac{1}{2}\p_\xi+c\right)\p_\sigma\left(\rho^2+c\rho\,\p_\xi\rho\right)\right]+\mathcal{O}(\epsilon).
 \la{eq:v1(rho)}
\end{align}

\subsection{Unstable weakly non-linear dynamics of the Kelvin mode}

Since Eq.~(\ref{eq:rho-edge}) imposes that $\varrho\sim \mathcal O(\epsilon)$, to determine the weakly nonlinear dynamics for the Kelvin mode, we only need to impose the no-penetration condition:
\begin{align}
    v_1\Big|_{\xi=0}=0\,.
\end{align}

Using the expression of $v_1$ given in Eq.~\ref{eq:v1(rho)} we can rewrite the no-penetration condition as a dynamical equation for the derivative of the edge density:
\begin{align}
    v_1\Big|_{\xi=0}=\frac{1}{2}\left[\p_\tau\p_\xi\rho-\frac{c}{4c^2+1}\p_\sigma\Big((\p_\xi\rho)^2\Big)\right]\bigg|_{\xi=0}=0\,.
\end{align}
In terms of $\eta(\tau,\sigma)$, this equation becomes he inviscid Burgers' equation:
\begin{equation}
   \p_\tau\eta-\frac{2c^2 \sqrt{2-c^2}}{4c^2+1}\eta\,\p_\sigma\eta+\mathcal{O}(\epsilon)=0\,, 
\end{equation}

Solutions of this equation are known to possess a finite time singularity. This means that the Kelvin mode is unstable and derivatives of the density grow indefinitely. One might ask whether this implies that the scaling used in these notes is not suitable to study the Kelvin mode, however, from the linear analysis we see that this mode does not have any dispersive corrections to regulate the growth of $\partial_\sigma \eta$. The instability of the near-edge dynamics for the Kelvin mode is consistent with Boyd's work on equatorial Kelvin waves, where the instability is mitigated by dispersive corrections arising from shearing motion. However, to the best of our knowledge, there is no additional mechanism to resolve this instability in the case of the Laughlin state, in the context of the CSGL theory. Whether such an instability can manifest in a quantum system remains an open question, which we do not address further in this work and leave for future investigation.

\section{Experimental Connections}

In this section, we will discuss the experimental observability of the weakly non-linear KdV edge dynamics.  Two possible experimental avenues can verify some of the predictions made in this paper. The first one is the well-established Gallium Arsenide heterostructures where there has been experimental evidence of non-linear shock wave propagation at the edge and a non-linear dissipative scaling at the edge. More recently, an FQH state with tunable edge confinement that could be atomically sharp has been realized in a monolayer graphene device fitted with an electrostatically tunable quantum point contact~\cite{cohen2024spontaneous}. These platforms enable enhanced experimental control over the boundary confinement of an FQH state and can potentially probe the non-linear edge dynamics. More recent observations of an Anomalous FQH state (without magnetic field) in a new graphene moir\'e system formed by pentalayer rhombohedral graphene and hBN~\cite{lu2024fractional} could be a promising platform to probe the edge dynamics in particular due to the absence of an external magnetic field. Carrying out the old GaAs experiments~\cite{talyanskii1994experimental, zhitenev1995linear} that reported evidence of non-linear effects within these modern Graphene FQH devices could shed more insights into the non-linear aspects of the edge dynamics.

A second platform to realizing the physics discussed in this paper involves synthetic quantum matter systems, particularly trapped atomic gases. In these systems, rapid rotation of the atomic cloud can induce synthetic magnetic fields, which are crucial for forming the many-body state in the fractional quantum Hall (FQH) regime. Although there are several challenges, this approach has seen some preliminary successes \cite{gemelke2010rotating, fletcher2021geometric}. The non-linear dynamics in these systems can be directly probed by experimentally preparing an initial state and observing its time evolution. The inherent non-linearities in these systems may lead to competing dispersive and non-linear effects at the edge, potentially resulting in soliton formation, provided the experimental time scales can be extended to weakly non-linear regimes where stable KdV solitons can emerge. A combination of numerical simulations and multiscale analysis is needed to estimate realistic time scales for soliton formation in a specific experimental setup, which we intend to explore in future works.

The KdV equation we derived has a specific testable prediction that can be verified through numerical calculations. Nardin and Carusotto in Ref.~\cite{nardin2023linear} determined the dispersive part of the KdV dynamics for the Laughlin state. The leading two terms in the dispersion of the KdV equation~(\ref{eq:KdV}) are $\omega\approx2 c k-\tfrac{1}{2}c \ell_B^2k^3$. The full dispersion relation is parameterized by a single non-universal interaction parameter $c$. However,  the ratio of the coefficient of linear dispersion to the coefficient of the cubic term only depends on the magnetic length.
\begin{align}
    \omega'(k\rightarrow 0)/\omega'''(k \rightarrow 0)=4 \ell_B^2.
\end{align}
Typically, this relation may depend on the details of the confining potential in a numerical setup. However, in the hard wall limit of the confining potential, this ratio could approach the value predicted by our theory.

\section{Generalizations and Future Directions}

In conclusion, we used the method of multiple scales to derive the Korteweg-de Vries (KdV) equation as the weakly nonlinear edge dynamics of composite boson hydrodynamics with hard wall boundary conditions. We employed the Hamiltonian framework for this KdV equation and demonstrated that the chiral Luttinger liquid theory can be recovered in the linearized regime. 

An interesting direction for future research is to use quantum KdV theory as a new starting point for studying the edge properties of the Laughlin state. This approach could also be extended to investigate edge transport in point contact setups by identifying the fermion operator within this Hilbert space that could act as the tunneling operator at the point contact.

In addition, this work can be easily extended to account for a constant external electric field and energy dissipation at the sample edge. As demonstrated in Refs.\cite{monteiro2022topological, monteiro2023coastal}, the presence of an external electric field not only modifies the Euler equation(\ref{Euler}) by introducing an external electric force to the fluid but also replaces the no-penetration condition~(\ref{no-penetration}) with the anomaly inflow boundary condition:
\begin{equation}
    \left(n v_y+\frac{\nu e}{2\pi\hbar}E_x\right)\bigg|_{y=0}=0\,.
\end{equation}
Both the tangent electric field and the energy dissipation at the edge introduce stresses at the boundary as discussed in~Ref.~\cite{monteiro2022topological, monteiro2023coastal}. By combining both these effects, Eq.~(\ref{no-stress}) gets modified and becomes:
\begin{align}
    \left[\p_t\sqrt{n}+\p_x\left(\sqrt{n}\,v_x\right)-\Gamma^2\p_x^2\left(\sqrt{n}\, v_x\right)+\frac{\nu e}{4\pi\hbar}E_x\right]\bigg|_{y=0}=0\,.
\end{align}
The scaling presented in Table~\ref{tab:scaling} remains valid in this more general setup as long as $E_y\sim\mathcal O(\epsilon)$, $E_x\sim \mathcal O(\epsilon^{5/2})$ and $\Gamma^2\sim\mathcal O(\sqrt{\epsilon})$. Hence, in the particular case where $E_y=0$, $E_x=\frac{\hbar\omega_B}{e\ell_B}\, \epsilon^{5/2}\,E$ and $\Gamma^2=\frac{\ell_B}{c}\sqrt{\epsilon}\,D$, we can follow the same steps as developed in section~\ref{sec:CBmode} to find that the chiral boson dynamics is governed by:
\begin{tcolorbox}[colback=green2!5!white,colframe=green2,title={\bf Diffusive and Anomalous Corrections to the KdV Equation}]
\begin{align}
   \p_\tau\varrho+ \frac{c}{2}\p_\sigma^3\varrho \boldsymbol{- D\p_\sigma^2\varrho}+\frac{4c^2 - 1}{4c^2 + 1}\,c\varrho\,\p_\sigma\varrho \boldsymbol{+\sqrt{\frac{\nu}{8\pi}}E}+ \mathcal O(\epsilon)=0\,.
\end{align}
\end{tcolorbox}
The terms in bold are the dissipative and anomaly corrections to Eq.~(\ref{eq:KdV}). Such modifications to the KdV equation can pave the way to studying edge transport beyond the linear regime.

\section*{Acknowledgments}
SG benefited greatly from stimulating discussions with Prof. Ganapathy Murthy and Prof. Ajit Balaram. SG is supported by NSF CAREER Grant No. DMR-1944967. Part of this work was conducted at the Aspen Center for Physics, which is supported by National Science Foundation Grant PHY-1607611.

\appendix
\section*{Appendix: Boundary Conditions from Variational Principle} \la{app:vpbc}

The boundary conditions derived from charge and energy conservation laws in Sec.~\ref{sec:bc} can also be obtained using a variational principle. In the following, we will introduce an effective edge action that can be incorporated into the original CSGL variables. Varying this effective edge action will produce the appropriate hydrodynamic boundary conditions discussed earlier. The equations of motion~(\ref{eq:theta} - \ref{eq:a}) emerge as saddle points of the action~(\ref{eq:csglaction}). This means that, when varying the action, the terms multiplying field variations in the bulk are set to zero, as they correspond to the equations of motion. By leaving the field variations at the boundary unconstrained, the boundary conditions are simply the equations of motion derived from the field variations projected onto the boundary. 

\section{No-penetration Condition} \label{sec:BC-action}

For a fluid confined to the lower half-plane, we observe that varying the action~(\ref{eq:csglaction}) generates the following boundary terms:
\begin{align}
    \delta S_{CSGL}&=\int dtdx \left[\frac{\hbar\nu}{4\pi}\left(\delta a_x\,a_0-\delta a_0\,a_x\right)-\frac{\hbar^2}{m}\delta\theta\,n\left(\partial_y\theta+a_y+\frac{e}{\hbar}A_y\right)-\frac{\hbar^2}{4m}\delta n\,\frac{\p_y n}{n}\right]_{y=0}. \label{eq:deltaS} 
    \end{align}

Here, we assume that the fields satisfy the bulk equations of motion. It is important to note that the variation~(\ref{eq:deltaS}) does not produce the boundary conditions discussed in the previous section. To derive these boundary conditions, we need to introduce a boundary action to be added to \(S_{\text{CSGL}}\).

The no-penetration condition for the original composite boson variables can be expressed as
\begin{align}
    \left(\partial_y\theta + a_y + \frac{e}{\hbar}A_y - \frac{\partial_x n}{2n}\right)\bigg|_{y=0} = 0\,. \label{eq:theta-edge}
\end{align}
Note that the variation \(\delta\theta\) only accounts for the first three terms, while the last term is not captured. Furthermore, to ensure that this boundary equation is consistent with one of the equations of motion (\ref{eq:a}), we must also impose the condition:
\begin{align}
   \left(\partial_t a_x - \partial_x a_0 + \frac{\pi \hbar}{\nu m} \partial_x n \right)\bigg|_{y=0} = 0\,. \label{eq:zeta}
\end{align}
However, this condition is equivalent to:
\begin{align}
    \left(a_0 - \frac{\pi \hbar}{\nu m} n \right)\bigg|_{y=0} &= \partial_t \zeta\,, \label{eq:ax-edge} \\
    a_x\bigg|_{y=0} &= \partial_x \zeta\,, \label{eq:a0-edge}
\end{align}
where \(\zeta\) is an arbitrary (undetermined) function.

The rationale is to obtain Eq.~(\ref{eq:theta-edge}) from the edge variation of $\theta$, Eq.~(\ref{eq:zeta}) from the variation of this new field $\zeta$ and Eqs.~(\ref{eq:ax-edge}-\ref{eq:a0-edge}) from the boundary variation of the gauge field components $a_x$ and $a_0$, respectively. It is straightforward to show that they can be derived by adding the following boundary action to the system:
\begin{align}
  S_{\text{edge}}&=-\int \left[\frac{\hbar^2}{2m}n\left(\partial_x\theta+\frac{1}{2} a_x+\frac{e}{\hbar}A_x\right)+\frac{\hbar\nu}{4\pi}\zeta\left(\partial_t a_x-\partial_x a_0+\frac{\pi \hbar}{\nu m}\partial_x n\right)\right]_{y=0}dtdx\,.
\end{align}


The second hydrodynamic boundary condition is derived from the variation of $n$ taken at the boundary, which gives us
\begin{align}
    \left(\partial_x\theta+\frac{1}{2}a_x+\frac{e}{\hbar}A_x+\frac{1}{2}\partial_x\zeta-\frac{1}{2}\partial_y n\right)\bigg|_{y=0}=0\,.
\end{align}
This expression becomes the no-slip boundary condition upon applying equation~(\ref{eq:a0-edge}). The first term in the edge action, \(S_{\text{edge}}\), is somewhat concerning as it does not initially appear in the gauge-invariant form \(\partial_x \theta + a_x\). However, the gauge invariance of this action becomes evident after a field redefinition, specifically after making the following substitutions:
\begin{align}
a_0 = \frac{\pi \hbar}{\nu m} n + \tilde{a}_0 \,, \qquad
a_i = \tilde{a}_i\,, \qquad \zeta = \tilde{\zeta} + \theta\,.
\end{align} 
Thus, denoting the combined action as \(S_{\text{no-slip}} = S_{\text{CSGL}} + S_{\text{edge}}\), the resulting action takes the familiar form derived in our previous work~\cite{monteiro2022topological}:
 \begin{align}
    S_{\text{no-slip}}&=-\int \left[\hbar n\left(\partial_t\theta+\tilde a_0\right)+\frac{\hbar^2}{2m}n\left(\partial_i\theta+\tilde a_i+\frac{e}{\hbar}A_i\right)^2+\frac{\hbar^2}{8mn}(\partial_i n)^2+\tilde V(n)\right.\nonumber
    \\
     &+\left.\frac{\hbar e B}{2m}\left(n-\frac{\nu eB}{2\pi\hbar}\right)+\frac{\hbar^2}{2m}n\,\epsilon_{ij}\partial_i\tilde a_j+\frac{\hbar\nu}{4\pi}\epsilon_{\alpha\beta\gamma}(\tilde a_\alpha+\partial_\alpha\theta)\partial_\beta\tilde a_\gamma\right]d^3x \nonumber
     \\
     &+\int dt dx\left[\frac{\hbar\nu}{4\pi}\tilde\zeta\left(\partial_x\tilde a_0-\partial_t\tilde a_x\right)-\frac{\hbar^2}{2m}n\left(\partial_x\theta+ a_x+\frac{e}{\hbar}A_x\right)\right]_{y=0},
\end{align}
where we set the chemical potential $\mu$ to lie at the lowest Landau level energy, that is, $\mu=\frac{\hbar eB}{2m}$.


Note that the no-penetration condition arises from varying the \(\theta\), \(\tilde{a}_0\), \(\tilde a_x\) and \(\tilde{\zeta}\) fields at the boundary, while the no-slip condition results from variations in the condensate density at the edge. However, the no-stress boundary condition acts as an additional dynamical equation in disguise, requiring the introduction of an auxiliary field at the boundary, as demonstrated in Refs.~\cite{monteiro2022topological, abanov2019hydro}. This new dynamical equation can be interpreted as the equation of motion for this additional degree of freedom.

\section{No-Stress Boundary Condition and Chiral Boson Action}
As discussed in sec.~\ref{sec:bc}, the no-stress condition can be reformulated as an additional dynamical equation:
\begin{align}
    T_{yx}\Big|_{y=0} = -\frac{\hbar n}{2}\left(\partial_y v_y - \partial_x v_x\right)\Big|_{y=0} = \hbar \sqrt{n}\left[\partial_t \sqrt{n} + \partial_x\left(\sqrt{n} v_x\right)\right]\Big|_{y=0} = 0\,, \la{eq:no-stress-dyn}
\end{align}
This equation can be interpreted as an additional conservation law at the edge of the sample. In this context, \(\sqrt{n}|_{y=0}\) represents the edge density, and \((\sqrt{n} v_x)|_{y=0}\) corresponds to the edge current. In the following, we derive the edge continuity equation as the equation of motion for an auxiliary field \(\phi\). To preserve the no-penetration condition, the additional action must depend only on \(n\) and the auxiliary field \(\phi\). Assuming that the fields satisfy the bulk equations of motion along with the boundary conditions~(\ref{eq:theta-edge}-\ref{eq:a0-edge}), we find:
\begin{align}
    \delta S_{\text{no-slip}} = \ldots - \frac{\hbar^2}{2m} \int dt\, dx\, \left(\delta n\, v_x\right)\Big|_{y=0}.
\end{align}
Following Refs.~\cite{abanov2020hydrodynamics, monteiro2022topological}, we observe that the action for the auxiliary field must take the form:
\begin{align}
    S_{\text{CB}} &= \frac{\hbar}{2} \int dt\, dx\, \partial_t \phi\,\big(\partial_x \phi - 2\sqrt{n}\big)\Big|_{y=0}.
\end{align}
However, this is valid only in the absence of an external electric field, as we have implicitly assumed \(A_x(t, x, 0) = 0\). For the general case, refer to~\cite{monteiro2022topological}.

By combining \(S_{\text{no-slip}}\) with \(S_{\text{CB}}\), the variation of \(n\) at the boundary gives:
\begin{align}
    \sqrt{n} v_x\Big|_{y=0} = \partial_t \phi, \label{eq:n-edge}
\end{align}
while the equation of motion for \(\phi\) is given by
\begin{align}    
\left(\partial_t\sqrt{n}+\partial_x\partial_t\phi\right)\big|_{y=0}=0\, . \la{eq:phi}
\end{align}
By using Eq.~(\ref{eq:n-edge}), this expression matches Eq.~(\ref{eq:no-stress-dyn}). Additionally, Eq.~(\ref{eq:phi}) implies that~\footnote{This is true up to a function of \(x\), but this function can be absorbed through a redefinition of \(\phi\).}
\begin{equation}
    \sqrt{n}\Big|_{\xi=0} = -\partial_x \phi\,. \label{eq:bosonization}
\end{equation}

The expressions for \(\sqrt{n}\big|_{\xi=0}\) and \((\sqrt{n} v_x)\big|_{\xi=0}\) at the boundary resemble the bosonization expressions in chiral Luttinger liquid theory. This supports the earlier identification, where \(\sqrt{n}\big|_{\xi=0}\) corresponds to the edge density, and \((\sqrt{n} v_x)\big|_{\xi=0}\) represents the edge current.

We emphasize that the effective action we derive incorporates several features of the chiral boson edge theory from the Chern-Simons formalism. We claim that this provides a complete description of the Laughlin state, incorporating both bulk and boundary aspects. However, this action is classical, and a more quantitative comparison with the chiral boson theory would require quantizing this action. The direct quantization of this effective edge action is beyond the scope of this work and will be addressed in a separate publication. 

\bibliographystyle{my-refs}
\bibliography{oddviscosity-bibliography}


\end{document}